\documentclass[twocolumn]{article}

\input{tcilatex}

\begin{document}

\subsubsection{\textbf{Fundamentals and Applications of Isotope Effect in
Modern Technology.}}

\textbf{V.G. Plekhanov.}

\bigskip

\textit{Fonoriton Science Lab., Garon Ltd., P.O. Box 2632, Tallinn, 13802,
ESTONIA}

\TEXTsymbol{<}e-mail\TEXTsymbol{>} vgplekhanov@hotmail.com

\bigskip

Different crystals (semiconductors and insulators) with varying isotopic
composition have been recently grown. I discuss here the effect of isotopic
mass and isotopic disorder on the properties (vibrational, elastic, thermal
and optical) of different crystals. The main applications of the stable
isotopes are included self-diffusion, neutron transmutative doping (NTD) of
different semiconductors, optical fibers, isotope-based quantum computers,
etc. Because of space limitations this discussion will not exhaustive. I
hope however, to give sufficient references to published work so that the
interest reader can easily find the primary literature sources to this
rapidly expanding field of solid state physics.

\textbf{\bigskip }

\textbf{Phonons, excitons, isotope-mixed crystals, laser materials, quantum
information, isotope-based quantum computers. }

\bigskip

It is well-known that the presence of randomly distributed impurities in a
crystal can give rise to significant variations of its mechanical,
electrical, thermal, and optical properties with respect to those of the
pure solid. All these properties are, more or less, directly related to the
structure of the manifold of phonon states and any variation induced in this
structure by the presence of the impurities, will produce a corresponding
alteration of the physical properties of the material. Of particular
interest is the case in which the impurity species is of the same chemical
nature, but with a different mass, i.e. the case of isotopic impurities. The
mechanisms by which the impurities (isotopes) perturb the phonon
distribution will depend on the mass difference between the host and guest
species [1-3]. Phonons are the crystal excitations most directly related to
the isotopic masses. In monatomic crystals (like C, Ge, Si., etc.), and
within the harmonic approximation, all phonon frequencies scale like the
square root of the average isotopic mass. Namely, this feature can be used
for the nondestructive isotopic \ characterization investigated materials.
The isotopic effect can be classified into two categories: 1) The first type
is caused by the variation of the phonon frequencies with the average
isotopic mass. To this type belongs the isotope effect in superconductors,
which plays an important role in the search for the mechanism of high T$_{c}$
superconductivity (see, e.g. [4]). The effect of changing the atomic mass M
is to change the phonon frequencies $\omega $ according to:

$\omega $ = $\sqrt{\frac{\alpha }{\text{M}}}$, \ \ \ \ \ \ \ \ \ \ \ \ \ \ \
\ \ \ \ \ \ \ \ \ \ \ \ \ \ \ \ \ \ \ \ \ \ \ \ \ \ \ \ \ \ \ \ \ \ \ \ \ \
\ \ \ \ \ \ \ \ \ \ (1)

where $\alpha $ is a force constant characteristic of the phonon under
consideration. The change in atomic mass implies, at low temperatures (see
below), a change in the average atomic displacement for each phonon mode. In
the case of one atom per primitive cell the mean squared phonon amplitude $%
\langle $u$^{2}\rangle $ is given by [1;2]:

$\langle $u$^{2}\rangle $ = $\langle \frac{\hbar ^{2}}{\text{4M}\omega }%
\left[ 1\text{ + 2n}_{B}\text{(}\omega \text{)}\right] \rangle $ = $\langle 
\frac{\hbar }{\text{4M}^{1/2}\alpha ^{1/2}}\left[ 1\text{ + 2n}_{B}\text{(}%
\omega \text{)}\right] \rangle $, \ \ \ \ \ \ \ (2)

where n$_{B}$($\omega $) is the Bose - Einstein statistical factor, $\omega $
is the frequency of a given phonon and $\langle $...$\rangle $ represents an
average over all phonon modes. \ The average in r.h.s. of (2) is often
simplified by taking the value inside $\langle $...$\rangle $ at an average
frequency $\omega _{D}$ which usually turns out to be close to the Debye
frequency. We should distinguish between the low temperature ($\hbar \omega $
\TEXTsymbol{>}\TEXTsymbol{>} k$_{B}$T) and the high temperature ($\hbar
\omega $ \TEXTsymbol{<}\TEXTsymbol{<} k$_{B}$T) limits and see:

($\hbar \omega $ \TEXTsymbol{>}\TEXTsymbol{>} k$_{B}$T), \ \ \ \ \ \ \ \ \ \ 
$\langle $u$^{2}\rangle $ = $\frac{\hbar }{\text{4M}\omega _{D}}$ $\sim $ M$%
^{-1/2}$ \ \ independent of T and

($\hbar \omega $ \TEXTsymbol{<}\TEXTsymbol{<} k$_{B}$T), \ \ \ \ \ \ \ \ \ \ 
$\langle $u$^{2}\rangle $ = $\frac{\text{k}_{B}\text{T}}{\text{2M}\omega ^{2}%
}\sim $ T \ \ \ \ \ \ \ \ \ independent of M \ \ \ \ (3).

Using Eq. (1) we can find from last equations that $\langle $u$^{2}\rangle $%
, the zero-point vibrational amplitude, is proportional to M$^{-1/2}$ \ at
low temperatures: it thus decrease with increasing M and vanishes for M$%
\longrightarrow \infty $. For high T, however, we find that \ $\langle $u$%
^{2}\rangle $ is independent of M and linear in T (details see [3] and
references therein).

Another type of isotope effects is produced by the isotopic mass
fluctuations about the average mass $\langle $M$\rangle $. These
fluctuations perturb the translational invariance of a crystal and lift, at
least in part, \textbf{k} - vector conservation. The most striking effect of
this type is observed in the thermal conductivity which has a maximum at a
temperature T$_{M}$ \TEXTsymbol{<}\TEXTsymbol{<} $\Theta _{D}$(here $\Theta
_{D}$ is Debye temperature, T$_{M}$ = 80 K for diamond, T$_{M}$ = 20 K for
silicon (see, also Figs. 64 - 66 in [3]). Reduction of the concentration of $%
^{13}$C from the standard 1\% (against 99\% of $^{12}$C) by a factor of ten
increases the thermal conductivity of diamond by about a factor of two, a
fact that leads to amplifications in situations where a large amount of
generated heat has to be driven away (e.g. as substrates for high power
electronic devices [5]). As is well-known this maximum represents the
transition from boundary scattering to the phonon unklapp scattering regime
and its value K$_{m}$ is determined by the isotopic fluctuation parameter g
(mass variance):

g = $\frac{\left\langle \text{M}^{2}\right\rangle }{\left\langle
M\right\rangle ^{2}}$ - 1, \ \ \ \ \ \ \ \ \ \ \ \ \ \ \ \ \ \ \ \ \ \ \ \ \
\ \ \ \ \ \ \ \ \ \ \ \ \ \ \ \ \ \ \ \ \ \ \ \ \ \ \ \ \ \ \ \ \ \ \ \ \ \
\ \ \ \ \ \ \ \ \ \ \ \ \ \ (4)

the larger g - the smaller K$_{m}$ [6].

It is known that materials having a diamond structure are characterized by
the triply degenerate phonon states in the $\Gamma $ - point of the
Brillouin zone (\textbf{k} = 0). These phonons are active in the Raman
scattering (RS) spectra, but not in the IR absorption ones \ (see, e.g.
[7]). First - order Raman light - scattering spectrum in diamond crystals
includes one line with the maximum at $\omega _{LTO}$($\Gamma $) = 1332.5 cm$%
^{-1}$. In Fig. 1$^{a}$, \ the first-order scattering spectrum in diamond
crystals with different isotope concentration is shown [8]. As was shown,
the maximum and the width of the first-order scattering line in
isotopically-mixed diamond crystals are nonlinearly dependent on the
concentration of isotopes x \ (see also [7]). The maximum shift of this line
is 52.3 cm$^{-1}$, corresponding to the limiting values of x = 0 and x = 1.

Fig. 1$^{b}$ demonstrates the dependence of the shape and position of the
first-order line of optical phonons in germanium crystal on the isotope
composition at liquid nitrogen temperatures [9]. The coordinate of the
center of the scattering line is proportional to the square root of the
reduced mass of the unit cell, i.e. M$^{-1/2}$. It is precisely this
dependence that is expected in the harmonic approximation (details see [3]).
An additional frequency shift of the line is observed for the natural and
enriched germanium specimens and is equal, as shown in Refs. [7, 9] to 0.34$%
\pm $0.04 and 1.06$\pm $0.04 cm$^{-1}$, respectively (see also Fig. 7 in
Chap. 4 of Ref. [10]). Detailed calculation of the shape of the lines in RS
of semiconductors have been performed by Spitzer et al. [11]. In their paper
a quantitative agreement with the experimental data on diamond and germanium
has been obtained. Comparing the half-widths of the scattering lines in
first-order RS in diamond and germanium (see Fig. 1), it is easy to see that
the observed line broadening due to isotopic disorder in diamond is much
greater than that in germanium. The reason for this is that the \textbf{k} =
0 point is not the highest point in the diamond dispersion curve (see Fig. 10%
$^{b}$ in Ref. [7]), whereas in the case of germanium it is the highest
point [12]. This shift of the maximum from the $\Gamma $ - point (\textbf{k}
= 0) leads to a much larger density of states in the vicinity of $\omega
_{LTO}$ in comparison with the normal one calculated by the formula:

N$_{d}$ $\sim $ Re($\omega _{LTO}$ - $\omega $ + i$\left[ \frac{\Delta
\omega _{LTO}}{\text{2}}\right] $)$^{1/2}$ \ \ \ \ \ \ \ \ \ \ \ \ \ \ \ \ \
\ \ \ \ \ \ \ \ \ \ \ \ \ \ (5).

(for more details see Ref. [12]). The density of states in diamond is
asymmetric with respect to $\omega _{LTO}$, causing asymmetry in the shape
of the scattering line [7]. This asymmetry also leads to the asymmetric
concentration dependence of the half-width of the scattering line. As was
shown early (see, e.g. [3] and references therein), in the case of a weak
potential of isotopic scattering of phonons, their self-energy $\varepsilon
\left( \omega \right) $ does not depend on \textbf{q} (- phonon
quasiimpuls). This is precisely the situation observed for C and Ge. Indeed,
if we express the mass fluctuation $\Delta $M/$\overline{\text{M}}$ ($%
\overline{\text{M}}$ is the mean mass of all isotopes) in the form of the
variation of the phonon band width $\Delta \omega _{0}$ = 12 cm$^{-1}$ at 
\textbf{q} = 0 and compare it with the width of the band of optical phonons
in Ge equals to $\approx $ 100 cm$^{-1}$, we will see that the variations
very small. Under this conditions the localization of optical phonons in Ge
is naturally, absent, and as observed in experiment, they stay delocalized
(see below, however opposite case in LiH$_{x}$D$_{1-x}$ crystals). Moreover,
direct measurements of the phonon lifetime in Ge show that, in the case of
anharmonic decay, it is two orders of magnitude shorter than the lifetime
that is due to the additional scattering by isotopes, i.e. $\tau _{anharm}$
= $\tau _{disord}\cdot $ 10$^{-2}$[13]. Therefore, the contribution of
anharmonicity to the half-width of the first-order light scattering line in
Ge is two orders of magnitude greater than that caused by the isotopic
disorder in crystal lattice. In conclusion of this part of our report we
should mention that analogous structure of first-order RS and their
dependence on isotope composition has by now been observed many times, not
only in elementary Si and $\alpha $-Sn, but also in compound CuCl, CuBr,
ZnSe, GaN semiconductors (details see Ref. [3]).

In Fig. 2 (curve 1) the spectrum of second-order RS of light in pure LiD
crystal is shown [7]. In spite of the fact, according to the nomogram of
exciton states [14], the crystal studied should be considered to be pure,
its RS spectrum contains a clear high-frequency peak around 1850 cm$^{-1}$.
The observed peak does not have an analogue in RS \ of pure LiH (Fig. 2,
curve 4) and has already been observed earlier in the second-order RS and
has been interpreted (see [7] and references therein) as a local vibration
of the hydrogen in LiD crystals. Further we note that as the concentration \
grows further (x \TEXTsymbol{>} 0.15) one observes in the spectra a
decreasing intensity in the maximum of 2LO($\Gamma $) phonons in LiD crystal
with a simultaneous growth in intensity of the highest frequency peak in
mixed LiH$_{x}$D$_{1-x}$crystals (Fig. 2, curve 3). The origin of the last
one is in the renormalization of LO($\Gamma $) vibrations in mixed crystals
[7]. Comparison of the structure of RS spectra (curves 1 and 2 in Fig. 2)
allows us, therefore, to conclude that in the concentration range of 0.1 
\TEXTsymbol{<} x \TEXTsymbol{<} 0.45 the RS spectra simultaneously contain
peaks of the LO($\Gamma $) phonon of pure LiD and the LO($\Gamma $) phonon
of the mixed LiH$_{x}$D$_{1-x}$crystal. Thus, the second-order RS spectra of
LiH$_{x}$D$_{1-x}$crystals have one- and two-mode character for LO($\Gamma $%
) phonons, and also contain a contribution from the local excitation at
small values of x. Moreover, we should add that an additional structure in
RS spectra on the short-side of the 2LO($\Gamma $) peak (see Fig. 21 in Ref.
[7]) was observed relatively ago in mixed LiH$_{x}$D$_{1-x}$crystals \ and,
very recently, in isotopically mixed crystals of diamond, germanium and $%
\alpha $-Sn (details see [3, 11]). These effects caused by isotopic disorder
in the crystal lattice of isotopically mixed crystals [3]. The observation
of two-mode behavior of the LO($\Gamma $) phonons in RS spectra of LiH$_{x}$D%
$_{1-x}$crystals contradicts the prediction of the CPA [15], according to
which the width W of optical vibration band should be smaller than the
frequency \ shift ($\Delta $) of transverse optical phonon. However, as was
shown early (see, e.g. [7] and references therein) in LiH$_{x}$D$_{1-x}$
mixed crystals, the reverse inequality is valid, i.e. W \TEXTsymbol{>} $%
\left\vert \Delta \right\vert $. According [16], this discrepancy between
experimental results and theory based on CPA [15] is mainly explained by the
strong potential of scattering of phonons, caused by a large change in the
mass upon substitution of deuterium for hydrogen. Once more reason of the
discrepancy between theory and results of the experiment may be connected
with not taking into account in theory the change of the force-constant at
the isotope substitution of the smaller in size D by H ion. We should stress
once more that among the various possible isotope substitution, by far the
most important in vibrational spectroscopy is the substitution of hydrogen
by deuterium. As is well-known, in the limit of the Born-Oppenheimer
approximation the force-constant calculated at the minimum of the total
energy depends upon the electronic structure and not upon the mass of the
atoms. It is usually assumed that the theoretical values of the phonon
frequencies depend upon the force-constants determined at the minimum of the
adiabatic potential energy surface. This leads to a theoretical ratio $%
\omega \left( \text{H}\right) $/$\omega \left( \text{D }\right) $of the
phonon frequencies that always exceed the experimental data. Very often
anharmonicity has been proposed to be responsible for lower value of this
ratio. In isotope effect two different species of the same atom will have
different vibrational frequencies only because of the difference in isotopic
masses. The ratio p of the optical phonon frequencies for LiH and LiD
crystals is given in harmonic approximation by:

p = $\frac{\omega \left( \text{H}\right) }{\omega \left( \text{D }\right) }$
= $\sqrt{\frac{\text{M}\left( \text{LiD}\right) }{\text{M}\left( \text{LiH}%
\right) }}\simeq $ $\sqrt{\text{2 }}$ \ \ \ \ \ \ \ \ \ \ \ \ \ \ \ \ \ \ \
\ \ \ \ \ \ \ \ \ \ (6)

while the experimental value (which includes anharmonic effects) is 1.396 $%
\div $ 1.288 (see Table in Ref. [17]). In this Table there are the
experimental and theoretical values of p according to formula (6), as well
as the deviation $\delta $ = $\frac{\text{P}_{Theory}\text{ - p}_{\exp }}{%
\text{p}_{theory}}$ of these values from theoretical ones. Using the least
squares method it was found the empirical formula of ln($\delta $\%) $\sim $
f(ln[$\frac{\partial \text{E}}{\partial \text{M}}]$) which is depicted on
Fig.3. As can be seen the indicated dependence has in the first
approximation a linear character:

ln($\delta $\%) = -7.5 + 2ln($\frac{\partial \text{E}}{\partial \text{M}}$).
\ \ \ \ \ \ \ \ \ \ \ \ \ \ \ \ \ \ \ \ \ \ \ \ \ \ \ \ \ \ (7)

From the results of Fig. 3, it can be concluded that only hydrogen compounds
(and its isotope analog - deuterium) need to take into account the
force-constant changes in isotope effect. It is also seen that for
semiconductor compounds (on Fig. 3 - points, which is below of Ox line) the
isotope effect has only the changes of the isotope mass (details see [3, 7]).

The dependence of the band gap energy on isotopic composition (via mechanism
of electron-phonon interaction) has already been observed for insulators
(Fig. 4) and lowest (indirect - direct) gap of different semiconductors ([3]
and references therein). It has been shown to result primarily from the
effect of the average \ isotopic mass on the electron-phonon interaction,
with a smaller contribution from the change in lattice constant. It was the
first paper [19] where the exciton binding energy E$_{B}$ was found to
depend on the isotopic composition. It was shown further that this change in
E$_{B}$ was attributed to the exciton-phonon interaction (originally with LO
phonons) (see, also [3]). At present time such dependence of E$_{B}$ $\sim $
f(x) (x- isotope concentration) was found for different bound excitons in
semiconductors [20 - 21]. The simplest approximation, in which crystals of
mixed isotopic composition are treated as crystals of identical atoms having
the average isotopic mass is referred to as virtual crystal approximation
(VCA) [15]. Going beyond the VCA, in isotopically mixed crystals one would
also expect local fluctuations in the band-gap energy from statistical
fluctuations in local isotopic composition within some effective volume,
such as that of an exciton (see, e.g. Fig. 2 of Ref. [18]). Using the
least-squares method it was found the empirical dependence of ln$\left( 
\frac{\partial \text{E}_{g}}{\partial \text{M}}\right) $ $\sim $ f$\left( 
\text{lnE}_{g}\right) ,$ which is presented on Fig. 5. As can be seen the
mentioned dependence has a parabolic character:

ln$\left( \frac{\partial \text{E}_{g}}{\partial \text{M}}\right) $ = 6.105$%
\left( \text{lnE}_{g}\right) ^{2}$ - 7.870$\left( \text{lnE}_{g}\right) $ +
0.565. \ \ \ \ \ \ \ \ (8)

From this figure it can be concluded also that the small variation of the
nuclear mass (semiconductors) causes the small changes in E$_{g}$ also. When
the nuclear mass increases \ it causes the large changes in E$_{g}$ (C, LiH,
CsH, etc.) (details, see [18, 3]).

Detail analyze the process of self-diffusion in isotope pure materials and
hetero-structures was done in [5]. Interest in diffusion in solids is as old
as metallurgy or ceramics, but the scientific study of the phenomenon may
probably be dated some sixth-seven decades ago. As is well-known, the
measured diffusion coefficients depends on the chemistry and structure of
the sample on which it is measured. In cited paper [5] it was shown to use
the stable isotopes for the study of diffusion process in different
semiconducting structures (bulk, hetero-structures etc.).

Chapter 6 indicated book [5] describes the new reactor technology - neutron
transmutative doping (NTD). Capture of thermal neutrons by isotope nuclei
followed by nuclear decay produces new elements, resulting in a very number
of possibilities for isotope selective doping of solids. The importance of
NTD technology for studies of the semiconductor doping as well as
metal-insulator transitions and neutral impurity scattering process is
underlined. The low-temperature mobility of free carriers in semiconductors
is mainly determined by ionized- and neutral-impurity scattering. The
ionized-impurity scattering mechanism has been extensively studied (see e.g.
[5] and references therein), and various aspects of this process are now
quite well understood. Scattering by neutral impurities is much less than by
ionized centers, i.e., its contribution is significant only in crystals with
low compensation and at very low temperatures where most of the free
carriers are frozen on the impurity sites. The availability of highly
enriched isotopes of Ge which can be purified to residual dopant levels 
\TEXTsymbol{<} 10$^{12}$ cm$^{-3}$ has provided the first opportunity to
measure neutral impurity scattering over a wide temperature range. In paper
[22] three Ge isotopes transmute into shallow acceptors (Ga), shallow donors
(As) and double donors (Se) (see also above):

$_{32}^{70}$Ge + n $\rightarrow $ $_{32}^{71}$Ge$_{EC(t_{1/2}=11.2days)}%
\rightarrow $ $_{32}^{71}$Ga + $\nu _{e},$

$_{32}^{74}$Ge + n $\rightarrow _{32}^{75}$Ge $_{\beta ^{-}(t_{1/2}\text{ = }%
82.2min)}$ $\rightarrow $ $_{32}^{75}$As + $\beta ^{-}$ + $\bar \nu _e$,

$_{32}^{76}$Ge + n $\rightarrow $ $_{32}^{77}$Ge$_{\beta _{{}}^{-}\text{(t}%
_{1/2}}$ $_{=}$ $_{11.3}$ $_{h\text{) }}\rightarrow $ $\beta ^{-}$ + $\bar{%
\nu}_{e}$ + $_{32}^{77}$As $_{\beta ^{-}\text{(t}_{1/2}}$ $_{=}$ $_{38.8}$ $%
_{h\text{)}}$ $\rightarrow $ $_{32}^{77}$Se + $\beta ^{-}$ + $\bar{\nu}_{e}$%
. (9)

The isotopes $^{72}$Ge and $^{73}$Ge are transmuted into the stable $^{73}$%
Ge and $^{74}$Ge respectively. Controlling the ratio of $^{70}$Ge and $^{74}$%
Ge in bulk Ge crystals allows fine tuning of the majority- as well as the
minority carrier concentration. Currently, this is the best method to vary
the free-carrier concentration independently from compensation ratio. As
opposed to other doping methods, NTD yields a very homogeneous, perfectly
random distribution of the dopants down to the atomic levels [5]. Thus
isotopically controlled crystals offer a unique possibility to study
systematically the scattering mechanism of the charge carriers in
semiconductors. Extensive Hall-effect and resistivity measurements from room
temperature down to 4.2K yielded very accurate free-carrier concentrations
and mobilities as a function of temperature and doping level were done in
paper [5]. Itoh et al. [22] have performed temperature-dependent Hall
measurements on four different p-type and two-different n-type Ge crystals
(Fig. 6). Fig. 6 shows the relative strength of the scattering from the
ionized and the neutral impurities. There is only a relatively small
temperature region in which the scattering from the neutral impurities
dominates. This range extends to higher temperatures as the free-carrier
concentration is increased. The calculated \textquotedblright transition
temperatures\textquotedblright\ above which the ionized impurities are the
main scattering centres compare very well with experimental results of Itoh
et al [22] (see also Fig. 6.31 in Ref. [5]). In order to demonstrate the
importance of the homogeneous dopant distribution, Itoh et al. have
performed the same study on samples cut from Ge : Ga crystals grown by the
conventional Czochralski method, where Ga impurities were introduced to Ge
melt during the crystal growth. These authors observed deviations of the
measured mobility from the theoretical calculations, which are most likely
due to inhomogeneous Ga impurity distributions in melt-doped Ge. Only the
use of NTD semiconductors with randomly distributed dopants allows for an
accurate test of the neutral impurity-scattering models (details, see [5]).

Another application of isotope pure and isotope mixed crystals that will be
discussed here is related to the possibility of using an isotopically mixed
medium (e.g. LiH$_{x}$D$_{1-x}$ or $^{12}$C$_{x}$ $^{13}$C$_{1-x}$) as an
oscillator of coherent radiation in the ultraviolet spectral range. To
achieve this, the use of indirect electron transitions involving, say, LO
phonons was planned [23]. The detection of LO phonon replicas of free -
exciton luminescence in wide - gap insulators attracted considerable
attention to these crystals (see e.g. [10; 23]). At the same time it is
allowed one to pose a question about the possibility of obtaining stimulated
emission in UV (VUV) region (4 - 6 eV) of the spectrum, where no solid state
sources for coherent radiation exist yet. In the first place this related to
the emitters working on the transitions of the intrinsic electronic
excitation (exciton). The last one provides the high energetical yield of
the coherent emission per unit volume of the substance.

In this part we will discuss the investigation results of the influence of
the excitation light density on the resonant secondary emission spectra of
the free - exciton in the wide - gap insulator LiH$_{x}$D$_{1-x}$ (LiH$%
_{1-x} $F$_{x}$) crystals. The cubic LiH crystals are typical wide - gap
ionic insulator with E$_{g}$ = 4.992 eV [10] with relatively weak exciton -
phonon interaction however: E$_{B}$/$\hbar \omega _{LO}$ = 0.29 where E$_{B%
\text{ }} $and $\hbar \omega _{LO}$ are exciton binding energy and
longitudinal optical phonon's energy , respectively. Besides it might be
pointed out that the analogous relation for CdS, diamond and NaI is 0.73;
0.45 and 12.7, respectively . In the insert of Fig. 7 depicts the
luminescence of 1LO and 2LO phonon replicas in LiH crystals. An increase in
the density of the exciting light causes a burst of the radiation energy in
the long-wave wing of the emission of the 1LO and 2LO repetitions (see Fig.
7) at a rate is higher for the 1LO replica line [23]. A detailed dependence
of the luminescence intensity and the shape of the 2LO phonon replica line
are presented in Fig. 7. The further investigations have shown [5] that with
the increase of the excitation light intensity at the beginning a certain
narrowing can be observed, followed by widening of the line of 2LO phonon
replica with a simultaneous appearance of a characteristics, probably mode
structure (see Fig. 8.11 in Ref. [5]). From \ this Fig. it can be seen that
the coupling between longwavelength luminescence intensity and excitation
light intensity is not only linear, but, in fact, of a threshold character
as in case of other crystals . A proximity of the exciton parameters of LiH
and CdS (ZnO) crystals allowed to carry out the interpretation of the
density effects in LiH on the analogy with these semiconducting compounds.
Coming from this in the paper [23] it was shown that for the observed
experimental picture on LiH crystals to suppose the exciton-phonon mechanism
of light generation [5] is enough the excitons density about 10$^{15}$ cm$%
^{-3}$. This is reasonable value, if the high quality of the resonator
mirrow - the crystal cleavage \textquotedblright in situ\textquotedblright\
and relatively large exciton radius (r = 40 \AA\ [10] )is taken into
account. To this light mechanism generation must be also promoting a large
value of the LO phonon energy $\left( \hbar \omega _{LO}\text{ = 140 meV }%
\right) .$ Owing to this the radiative transition is being realized in the
spectral region with a small value of the absorption coefficient, and thus
with a small losses in resonator (details see [5]).

In conclusion of this section we should underlined that if the observable
mode structure is really caused by the laser generation it may be smoothly
tuned in the region of energies 4.5 $\pm $ 5.1 eV owing to smooth transition
of the line emission energy in the LiH$_{x}$D$_{1-x}$ (LiH$_{x}$F$_{1-x}$;
LiD$_{x}$F$_{1-x}$) mixed crystals as well as in the range 5.35 - 5.10 eV in 
$^{12}$C$_{x}$ $^{13}$C$_{1-x}$ mixed crystals (see also [10]).

Concluding our report we should \ be paid your attention to the reports of
Professors Schoven, Weston, Wendt as well as Dr. Chai of our conference
which are devoted in the first step of radioactive isotope applications.

\bigskip

Figure Captions.

Fig. 1. a) First-order Raman spectra of $^{12}$C$_{x}^{13}$C$_{1-x}$
diamonds with different isotope compositions. The labels A,B, C, D, E and F
correspond to x = 0.989; 0.90; 0.60; 0.50; 0.30 and 0.01 respectively. The
intensity is normalized at each peak (after [8]); b) First-order Raman
scattering spectra in Ge with different isotope contents (after [13]).

Fig. 2. Second-order Raman spectra of LiH$_{x}$D$_{1-x}$ crystals at room
temperature: (1); (2); (3) and (4) x = 0; 0.42; 0.76 and 1, respectively
(after [7]).

Fig. 3. The dependence of ln($\delta $\%) $\sim $ f(ln[$\frac{\partial \text{%
E}}{\partial \text{M}}]$): points are experimental values and continuous
line - calculation on the formula (7) (after [17]).

Fig. 4. Mirror reflection spectra of crystals: LiH, curve 1; LiH$_{x}$D$%
_{1-x}$, curve 2 and LiD, curve 3 at 4.2 K. Light source without crystals,
curve 4 (after [18]).

Fig. 5. The dependence of ln$\left( \frac{\partial \text{E}_{g}}{\partial 
\text{M}}\right) $ $\sim $ f$\left( \text{lnE}_{g}\right) $: points are
experimental date and continuous line - calculation on the formula (8)
(after [18]).

Fig. 6. Temperature dependence of the carrier mobility of a) p - type and b)
n - type NTD Ge crystals. c) Temperature dependence of relative
contributions to the mobility. Note that the mobility is dominated by
neutral impurity scattering below 20 K ($^{70}$Ge:Ga $\sharp $ 2 crystal)
(after [22]).

Fig. 7. The dependence of the intensity in the maximum (1) and on the
long-wavelength side (2) of 2LO replica emission line of LiH crystals on the
excitation light intensity. In insert: luminescence spectra of free excitons
in LiH crystals in the region of the emission lines of 1LO and 2LO phonon
repetitions at 4.2 K for low (1) and high (2) density of excitations of 4.99
eV photons (after [23]).

\bigskip

References.

1. I.M. Lifshitz,\textit{\ Physics of Real Crystals and Disordered Systems},
Selected Works (Eds. M.I. Kaganov, A.M. Kosevich, Science, Moscow, 1987) (in
Russian).

2. A.A. Maradudin, E.W. Montroll, G.H. Weiss and I.P. Ipatova, \textit{%
Theory of Lattice Dynamics in the Harmonic Approximation}, Solid State
Physics, Vol.3, (Eds. F. Seitz, D. Turnbull and H. Ehrenreich, Academic, New
York, 1971).

3. V.G. Plekhanov, Elementary Excitations in Isotope-Mixed Crystals, \textit{%
Physics Reports}, \textbf{410} [1-3] 1 (2005).

4. J.P. Franck, in:\textit{\ Physical Properties of High T}$_{c}$\textit{\
Superconductors} (ed. D.M. Ginsberg, Vol. 4., World Scientific, Singapore,
1984) p. 189.

5. For a review, see, V.G. Plekhanov,\textit{\ Applications of the Isotopic
Effect in Solids}, Springer, Berlin - Heidelberg, 2004.

6. See, for example, R. Berman, \textit{Thermal Conduction of Solids}
(Clarendon Press, Oxford, 1976); T.M. Tritt, \textit{Thermal Conductivity}
(Springer, Berlin - Heidelberg, 2005).

7. V.G. Plekhanov, Isotopic Effects in Lattice Dynamics, \textit{Physics -
Uspekhi} (Moscow) 46 [7] 689 (2003).

8. H. Hanzawa, N. Umemura, Y. Nisida et al., Disorder Effects of Nitrogen
Impurities, Irradiation - Induced Defects, and $^{13}$C Isotope Composition
on the Raman Spectrum in Synthetic I$^{b}$ Diamond, \textit{Phys. Rev.} 
\textbf{B54} [6] 3793 (1996).

9.M. Cardona, Semiconductor Crystals with Tailor - Made Isotopic
Compositions, in: \textit{Festkorperprobleme/Advances in Solid State Physics}
(ed. R. Helbig, Vieweg, Braunschweig, Wiesbaden, Vol. 34, 1994) p. 35.

10. V.G. Plekhanov, \textit{Isotope Effects in Solid State Physics}
(Academic, New York, 2001).

11. J. Spitzer, P. Etchegoin, M. Cardona et al., Isotopic - Disorder Induced
Raman Scattering in Diamond, \textit{Solid State Commun. }\textbf{88} [6]
509 (1993).

12. M. Cardona, Isotopic Effects in the Phonon and Electron Dispersion
Relations of Crystals, \textit{Phys. Stat. Sol. }(b) \textbf{220} [1] 5
(2000).

13. M. Cardona, P. Etchegoin, H.D. Fuchs et al., Effect of Isotopic Disorder
and Mass on the Electronic and Vibronic Properties of Three-, Two- and One -
Dimensional Solids, \textit{J. Phys.: Condens. Matter} \textbf{5} [1] A61
(1993).

14. V.G. Plekhanov, Phonon renormalization of Interband Transition Energy in
the Mixed Crystals, \textit{Solid State Commun.} \textbf{76} [1] 51 (1990).

15. R.J. Elliott, J.A. Krumhansl and P.L. Leath, The Theory and Properties
of Randomly Disordered Crystals and Related Physical Systems,\textit{\ Rev.
Mod. Phys.} \textbf{46} [3] 465 (1974).

16. V.G. Plekhanov, Experimental Evidence of Strong Phonon Scattering in
Isotope Disordered Systems: The Case of LiH$_{x}$D$_{1-x}$, \textit{Phys.
Rev.} \textbf{B51} [14] 8874 (1995).

17. V.G. Plekhanov (unpublished results, 2004)

18. V.G. Plekhanov, N.V. Plekhanov, Isotope Dependence of Band - Gap Energy, 
\textit{Phys. Lett.}, \textbf{A313} [3] 231 (2003).

19. V.G. Plekhanov, T.A. Betenekova, V.A. Pustovarov, Excitons and
Peculiarities of Exciton-Phonon Interaction in LiH and LiD, \textit{Sov.
Phys. Solid State} \textbf{18} [8] 1422 (1976).

20. M. Cardona, Dependence of the Excitation Energies of Boron in Diamond on
Isotopic Mass, \textit{Solid State Commun. }\textbf{121} [1] 7 (2002).

21. D. Karaiskaj, T.A. Meyer, M.L.W. Thewalt et al., Dependence of the
Ionization Energy of Shallow Donors and Acceptors in Silicon on the Host
Isotopic Mass, \textit{Phys. Rev.}\textbf{\ B68} [2] 121201 (2003).

22. H.D. Fuchs, K.M. Itoh and E.E. Haller, Isotopically Controlled
Germanium: A New Medium for the Study of Carrier Scattering by Neutral
Impurities, \textit{Philos. Mag.} \textbf{B70} [2] 662 (1994); K.M. Itoh,
E.E. Haller, V.I. Ozogin, Neutral - Impurity Scattering in Isotopically
Engineering Ge, \textit{Phys. Rev.} \textbf{B50} [23] 16995 (1994).

23. V.G. Plekhanov, Wide-Gap Insulators Excitonic Nonlinearity and Its
Potential Applications in Solid State Lasers, in: \textit{Proc. Int. Conf.
Advances Solid State Lasers}, USA, SOQUE, 1990.

\textbf{Table. }Values of the coefficients dE/dM (mev, cm$^{-1}$) for the
optical phonons and \ the experimental and theoretical values of p as well
as deviation $\delta $\% of these values from theoretical ones.

\begin{tabular}{lllll}
Substances & Frequencies & p$_{\exp }$ & p$_{\text{theory}}$ & $\delta $\% = 
$\frac{\text{p}_{\text{theory}}\text{ - p}_{\text{exp}}}{\text{p}_{\text{%
theory}}}$ \\ 
LiH/LiD & 140(meV)/104(meV)[10,24] & 1.288-1.346 & $\sqrt{\text{2}}$ = 1.414
& 4.8 - 8.9 \\ 
SiH$_{4}$/SiD$_{\text{4}}$ & 2186.87/1563.3(cm$^{-1}$)[10] & 1.399 & $\sqrt{%
\text{2}}$ = 1.414 & 1.5 \\ 
$^{12}$C/$^{13}$C & 1332,5/1280(cm$^{-1}$)[10;25] & 1.041 & $\sqrt{\frac{%
\text{13}}{\text{12}}}$ = 1.041 & 0.001 \\ 
$^{70}$Ge/$^{76}$Ge & 309.8/297.7(cm$^{-1}$)[10,26,27] & 1.041 & $\sqrt{%
\frac{\text{76}}{\text{70}}}$ = 1.042 & 0.096 \\ 
$^{28}$Si/$^{30}$Si & 524.8/509.8(cm$^{-1}$)[28] & 1.029 & $\sqrt{\frac{%
\text{30}}{\text{28}}}$ = 1.035 & 0.58 \\ 
$^{64}$Zn$^{76}$Se/$^{68}$Zn$^{80}$Se & 213.2/207.4(cm$^{-1}$)[10] & 1.028 & 
$\sqrt{\frac{\mu _{1}}{\mu _{2}}}$ = 1.029 & 0.097 \\ 
$\alpha -^{112}$Sn/$\alpha $- $^{124}$Sn & 206.5/196.8(cm$^{-1}$)[10,30] & 
1.049 & $\sqrt{\frac{\text{124}}{\text{112}}}$ = 1.052 & 0.30 \\ 
Ga$^{14}$N/Ga$^{15}$N & 535/518(cm$^{-1}$)[10,31] & 1.033 & $\sqrt{\frac{%
\text{15}}{\text{14}}}$ = 1.035 & 0.19 \\ 
$^{63}$Cu$^{35}$Cl/$^{65}$Cu$^{37}$Cl & 174.4/171.6(cm$^{-1}$)[10,32] & 1.016
& $\sqrt{\frac{\mu _{1}}{\mu _{2}}}$ = 1.022 & 0.59%
\end{tabular}

\end{document}